\documentclass[12pt,letterpaper]{article}
\usepackage{epsfig,rotating,setspace,latexsym,amsmath,epsf,amssymb,bm}
\usepackage{cite}

\title{A New Achievability Scheme for the Relay Channel\thanks{This work was
supported by NSF Grants CCR $03$-$11311$, CCF $04$-$47613$ and CCF
$05$-$14846$, and was presented in part at IEEE Information Theory Workshop, Lake Tahoe, CA, September 2007.}}

\author{Wei Kang \qquad Sennur Ulukus \\
\normalsize Department of Electrical and Computer Engineering\\
\normalsize University of Maryland, College Park, MD 20742 \\
\normalsize {\it wkang@umd.edu} \qquad {\it ulukus@umd.edu}}

\newtheorem{Theo}{Theorem}

\begin{document}
%\date{}
\maketitle 

\begin{abstract}
In this paper, we propose a new coding scheme for the general relay
channel. This coding scheme is in the form of a block Markov code. The
transmitter uses a superposition Markov code. The relay compresses the
received signal and maps the compressed version of the received signal
into a codeword conditioned on the codeword of the previous block. The
receiver performs joint decoding after it has received all of the $B$
blocks. We show that this coding scheme can be viewed as a
generalization of the well-known Compress-And-Forward (CAF) scheme
proposed by Cover and El Gamal. Our coding scheme provides options for
preserving the correlation between the channel inputs of the
transmitter and the relay, which is not possible in the CAF
scheme. Thus, our proposed scheme may potentially yield a larger
achievable rate than the CAF scheme.
\end{abstract}

\newpage

\section{Introduction}
As the simplest model for cooperative communications, relay channel
has attracted plenty of attention since 1971, when it was first
introduced by van der Meulen \cite{Van_Der_Meulen:1971}. In 1979,
Cover and El Gamal proposed two major coding schemes for the relay
channel \cite{Cover:1979}. These two schemes are widely known as
Decode-And-Forward (DAF) and Compress-And-Forward (CAF) today; see
\cite{Kramer:2005} for a recent review. These two coding schemes
represent two different types of cooperation. In DAF, the cooperation
is relatively obvious, where the relay decodes the message from the
transmitter, and the transmitter and the relay cooperatively transmit
the constructed common information to the receiver in the next
block. In CAF, the cooperation spirit is less easy to recognize, as
the message is sent by the transmitter only once. However, the relay
cooperates with the transmitter by compressing and sending its signal
to the receiver. The rate gains in these achievable schemes are due to
the fact that, through the channel from the transmitter to the relay,
{\it correlation} is created between the transmitter and the relay,
and this correlation is utilized to improve the rates.

In the DAF scheme, correlation is created and then utilized in a block
Markov coding structure. More specifically, a {\it full} correlation
is created by decoding the message fully at the relay, which enables
the transmitter and the relay to create any kind of joint distribution
for the channel inputs in the next block. %; in an additive Gaussian
%channel, for instance, they transmit with the {\it full correlation},
%i.e., ``beamform'' to the receiver, as this channel favors
%correlation. 
The shortcoming of the DAF scheme is that by forcing the
relay to decode the message in its entirety, it limits the overall
achievable rate by the rate from the transmitter to the relay. In
contrast, by not forcing a full decoding at the relay, the CAF scheme
does not limit the overall rate by the rate from the transmitter to
the relay, and may yield higher overall rates. The shortcoming of the
CAF scheme, on the other hand, is that the correlation offered by the
block coding structure is not utilized effectively, since in each
block the channel inputs $X$ and $X_1$ from the transmitter and the
relay are independent, as the transmitter sends the message only once.

However, the essence of good coding schemes in multi-user systems with
correlated sources (e.g., \cite{Cover:1980, Ahlswede:1983}) is to
preserve the correlation of the sources in the channel
inputs. Motivated by this basic observation, in this paper, we propose
a new coding scheme for the relay channel, that is based on the idea
of preserving the correlation in the channel inputs from the
transmitter and the relay. We will show that our new coding scheme may
be viewed as a more general version of the CAF scheme, and therefore,
our new coding scheme may potentially yield larger rates than the CAF
scheme. Our proposed scheme can be further combined with the DAF
scheme to yield rates that are potentially larger than those offered
by both DAF and CAF schemes, similar in spirit to \cite[Theorem
7]{Cover:1979}.

Our new achievability scheme for the relay channel may be viewed as a
variation of the coding scheme of Ahlswede and Han
\cite{Ahlswede:1983} for the multiple access channel with a correlated
helper. In our work, we view the relay as the helper because the
receiver does not need to decode the information sent by the
relay. Also, we note that the relay is a {\it correlated helper} as
the communication channel from the transmitter to the relay provides relay
for free a correlated version of the signal sent by the
transmitter. The key aspects of the Ahlswede-Han \cite{Ahlswede:1983}
scheme are: to preserve the correlation between the channel inputs of
the transmitter and the helper (relay), and for the receiver to decode
a ``virtual'' source, a compressed version of the helper, but not the
entire signal of the helper.

Our new coding scheme is in the form of block Markov coding. The
transmitter uses a superposition Markov code, similar to the one used
in the DAF scheme \cite{Cover:1979}, except in the random codebook
generation stage, a method similar to the one in \cite{Cover:1980} is
used in order to preserve the correlation between the blocks. Thus, in
each block, the fresh information message is mapped into a codeword
conditioned on the codeword of the previous block. Therefore, the overall codebook at the transmitter has a tree structure, where the codewords in block $l$  emanate from the codewords in block $l-1$. The depth of the tree is $B-1$. A similar strategy
is applied at the relay side where the compressed version of the
received signal is mapped into a two-block-long codeword conditioned
on the codeword of the previous block. Therefore, the overall codebook at the relay has a tree structure as well. As a result of this coding strategy, we successfully
preserve the correlation between the channel inputs of the transmitter
and the relay. However, unlike the DAF scheme where a {\it full}
correlation is acquired through decoding at the relay, our scheme
provides only a {\it partially} correlated helper at the relay by not
trying to decode the transmitter's signal fully. From
\cite{Cover:1980, Ahlswede:1983}, we note that the channel inputs are
correlated through the virtual sources in our case, and therefore, the
channel inputs between the consecutive blocks are correlated. This
correlation between the blocks will surely hurt the achievable
rate. The correlation between the blocks is the price we pay for
preserving the correlation between the channel inputs of the
transmitter and the relay within any given block.

At the decoding stage, we perform joint decoding for the entire $B$
blocks after all of the $B$ blocks have been received, which is
different compared with the DAF and CAF schemes. The reason for
performing joint decoding at the receiver is that due to the
correlation between the blocks, decoding at any time before the end of
all the $B$ blocks would decrease the achievable rate. We note that
joint decoding increases the decoding complexity and the delay as
compared to DAF and CAF, though neither of these is a major concern in
an information theoretic context. The only problem with the joint
decoding strategy is that it makes the analysis difficult as it
requires the evaluation of some mutual information expressions
involving the joint probability distributions of up to $B$ blocks of
codes, where $B$ is very large.

The analysis of the error events provides us three conditions
containing mutual information expressions involving infinite letters
of the underlying random process. Evaluation of these mutual
information expressions is very difficult, if not impossible. To
obtain a computable result, we lower bound these mutual informations
by noting some Markov structure in the underlying random process. This
operation gives us three conditions to be satisfied by the achievable
rates. These conditions involve eleven variables, the two channel
inputs from the transmitter and the relay, the two channel outputs at
the relay and the receiver and the compressed version of the channel
output at the relay, in two consecutive blocks, and the channel input
from the transmitter in the previous block.

We finish our analysis by revisiting the CAF scheme. We develop an
equivalent representation for the achievable rates given in
\cite{Cover:1979} for the CAF scheme. We then show that this
equivalent representation for the achievable rates for the CAF scheme
is a special case of the achievable rates in our new coding scheme,
which is obtained by a special selection of the eleven variables
mentioned above. We therefore conclude that our proposed coding scheme
yields potentially larger rates than the CAF scheme. More importantly,
our new coding scheme creates more possibilities, and therefore a
spectrum of new achievable schemes for the relay channel through the
selection of the underlying probability distribution, and yields the
well-known CAF scheme as a special case, corresponding to a particular
selection of the underlying probability distribution.

\section{The Relay Channel}
Consider a relay channel with finite input alphabets $\mathcal{X}$,
$\mathcal{X}_1$ and finite output alphabets $\mathcal{Y}$,
$\mathcal{Y}_1$, characterized by the transition probability
$p(y,y_1|x,x_1)$. An $n$-length block code for the relay channel
$p(y,y_1|x,x_1)$ consists of encoders $f, f_i$, $i=1,\dots,n$ and a
decoder $g$
\begin{align}
f&: \mathcal{M}\longrightarrow \mathcal{X}^n\nonumber\\
f_i&: \mathcal{Y}_1^{i-1}\longrightarrow \mathcal{X}_1, \qquad i=1,
\dots,n\nonumber\\
g&: \mathcal{Y}^n\longrightarrow \mathcal{M}\nonumber
\end{align}
where the encoder at the transmitter sends $x^n=f(m)$ into the
channel, where $m \in \mathcal{M}\triangleq \{1,2,\dots, M\}$; the encoder at the relay at the
$i$th channel instance sends $x_{1i}=f_i(y_1^{i-1})$ into the channel;
the decoder outputs $\hat{m}=g(y^n)$. The average probability of
error is defined as
\begin{equation}
P_e=\frac{1}{M}\sum_{m\in \mathcal{M}}Pr(\hat{m}\neq m|m 
~\text{is transmitted})
\end{equation}
A rate $R$ is achievable for the relay channel $p(y,y_1|x,x_1)$ if for
every $0<\epsilon<1$, $\eta>0$, and every sufficiently large $n$,
there exists an $n$-length block code $(f,f_i, g)$ with $P_e\le
\epsilon$ and $\frac{1}{n}\ln M\ge R-\eta$.

\section{A New Achievability Scheme for the Relay Channel}\label{SL1}
We adopt a block Markov coding scheme, similar to the DAF and CAF
schemes. We have overall $B$ blocks. In each block, we transmit
codewords of length $n$. We denote the variables in the $l$th block
with a subscript of $[l]$. We denote $n$-letter codewords transmitted
in each block with a superscript of $n$. Following the standard relay
channel literature, we denote the (random) signals transmitted by the
transmitter and the relay by $X$ and $X_1$, the signals received at
the receiver and the relay by $Y$ and $Y_1$, and the compressed
version of $Y_1$ at the relay by $\hat{Y}_1$. The realizations of
these random signals will be denoted by lower-case letters. For
example, the $n$-letter signals transmitted by the transmitter and the
relay in the $l$th block will be represented by $x_{[l]}^n$ and
$x_{1[l]}^n$.

Consider the following discrete time stationary Markov process
$G_{[l]}\triangleq(X,\hat{Y}_1,X_1,y, Y_1)_{[l]}$ for $l=0,1,\dots,B$,
with the transition probability distribution
\begin{align} \label{distr}
p\left((x,\hat{y}_1,x_1,y,y_1)_{[l]}|(x,\hat{y}_1,x_1,y,y_1)_{[l-1]}\right)&
\nonumber\\
=p(x_{[l]}|x_{[l-1]})p&(y_{1[l]},y_{[l]}|x_{[l]},x_{1[l]})p(x_{1[l]}|
\hat{y}_{1[l-1]})p(\hat{y}_{1[l]}|y_{1[l]}, x_{1[l]})
\end{align} 
The codebook generation and the encoding scheme for the $l$th block,
$l=1,\dots, B-1$, are as follows.

\begin{figure*}
\centering
\includegraphics[width=6.3in]{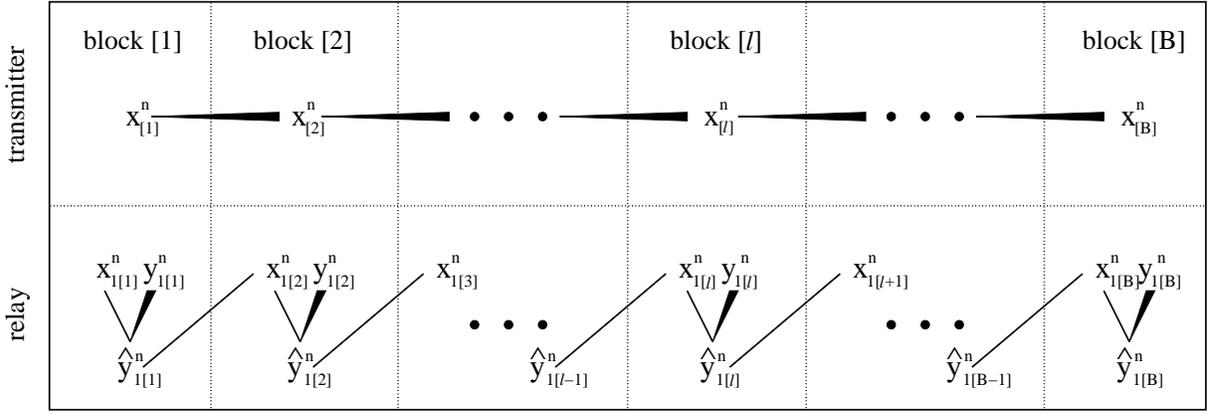}
\caption{Codebook structure.}
\label{codeS}
\end{figure*}

\vspace*{0.1in}
\noindent
{\bf Random codebook generation:} Let $(x_{[l-1]}^n(m_{[l-1]}),
x_{1[l-1]}^n,  y_{1[l-1]}^n, y_{[l-1]}^n)$ denote the
transmitted and the received signals in the $(l-1)$st block, where
$m_{[l-1]}$ is the message sent by the transmitter in the $(l-1)$st
block. An illustration of the codebook structure is shown in
Figure~\ref{codeS}.

\begin{enumerate}
\item For each $x_{[l-1]}^n(m_{[l-1]})$ sequence, generate $M$
sequences, where $x^n_{[l]}(m_{[l]})$, the $m_{[l]}$th sequence, is
generated independently according to $\prod_{i=1}^n
p(x_{i[l]}|x_{i[l-1]})$. Here, every codeword in the $(l-1)$st block
expands into a codebook in the $l$th block. This expansion is
indicated by a directed cone from $x_{[l-1]}^n$ to $x_{[l]}^n$ in
Figure~\ref{codeS}.

\item For each $x_{1[l-1]}^n$ sequence, generate $L$
$\hat{Y}_{1[l-1]}^n$ sequences independently uniformly distributed in
the conditional strong typical set\footnote{Strong typical set and
conditional strong typical set are defined in \cite[Definition 1.2.8,
1.2.9]{Csiszar:1981}. For the sake of simplicity, we omit the
subscript which is used to indicate the underlying distribution in
\cite{Csiszar:1981}.}  $\mathcal{T}_{\delta}(x_{1[l-1]}^n)$ with
respect to the distribution $p(\hat{y}_{1[l-1]}|x_{1[l-1]})$.  If
$\frac{1}{n}\ln L>I(Y_{1[l-1]};\hat{Y}_{1[l-1]}|X_{1[l-1]})$, for any
given $y_{1[l-1]}^n$ sequence, there exists one $\hat{y}_{1[l-1]}^n$
sequence with high probability when $n$ is sufficiently large such
that $(y_{1[l-1]}^n, \hat{y}_{1[l-1]}^n, x_{1[l-1]}^n)$ are jointly
typical according to the probability distribution $p(y_{1[l-1]},
\hat{y}_{1[l-1]}, x_{1[l-1]})$. Denote this $\hat{y}_{1[l-1]}^n$ as
$\hat{y}_{1[l-1]}^n(y_{1[l-1]}^n, x_{1[l-1]}^n)$. Here, the
quantization from $y_{1[l-1]}^n$ to $\hat{y}_{1[l-1]}^n$, parameterized
by $x_{1[l-1]}^n$, is indicated in Figure~\ref{codeS} by a directed
cone from $y_{1[l-1]}^n$ to $\hat{y}_{1[l-1]}^n$, with a straight line
from $x_{1[l-1]}^n$ for the parameterization.

\item For each $\hat{y}_{1[l-1]}^n$, generate one $x_{1[l]}^n$
sequence according to $\prod_{i=1}^n
p(x_{1i[l]}|\hat{y}_{1i[l-1]})$. This one-to-one mapping is indicated
by a straight line between $\hat{y}_{1[l-1]}^n$ and $x_{1[l]}^n$ in
Figure~\ref{codeS}.
\end{enumerate}

\vspace*{0.1in}
\noindent
{\bf Encoding:} Let $m_{[l]}$ be the message to be sent in this
block. If $(x_{[l-1]}^n(m_{[l-1]}), x_{1[l-1]}^n)$ are sent and $y_{1[l-1]}^n$ is received in the previous block,  
we choose   $(x^n_{[l]}(m_{[l]}), \hat{y}_{1[l-1]}^n(y_{1[l-1]}^n,
x_{1[l-1]}^n), x_{1[l]}^n)$ according to the code generation method
described above  and transmit $(x^n_{[l]}(m_{[l]}), x_{1[l]}^n)$. In the
first block, we assume a virtual $0$th block, where $(x^n_{[0]},
x^{n}_{1[0]}, \hat{y}^n_{1[0]})$, as well as $x^n_{1[1]}$, are known by the transmitter, the
relay and the receiver. In the $B$th block, the transmitter randomly
generates one $x_{[B]}^n$ sequence according to $\prod_{i=1}^n
p(x_{i[B]}|x_{i[B-1]})$ and sends it into the channel. The relay,
after receiving $y_{1[B]}^n$, randomly generates one
$\hat{y}_{1[B]}^n$ sequence according to $\prod_{i=1}^n
p(\hat{y}_{1i[B]}|y_{1i[B]},x_{1i[B]})$. We assume that the
transmitter and the relay reliably transmit $x_{[B]}^n$ and
$\hat{y}_{1[B]}^n$ to the receiver using the next $b$ blocks, where
$b$ is some finite positive integer. We note that $B+b$ blocks are
used in our scheme, while only the first $B-1$ blocks carry the
message. Thus, the final achievable rate is
$\frac{B-1}{B+b}\frac{1}{n}\ln M$ which converges to $\frac{1}{n}\ln
M$ for sufficiently large $B$ since $b$ is finite.

\vspace*{0.1in} 
\noindent
{\bf Decoding:} After receiving $B$ blocks of $y^n$ sequences, i.e.,
$y^n_{[1]},\dots, y^n_{[B]}$, and assuming $x^n_{1[1]}$, $x^n_{[B]}$ and
$\hat{y}_{1[B]}^n$ are known at the receiver, we seek
$x^n_{[1]},\dots, x^n_{[B-1]}$, $\hat{y}^n_{1[1]},\dots,
\hat{y}^n_{1[B-1]}, x^n_{1[2]},\dots, x^n_{1[B]}$, such that
\begin{equation}
\left(x^n_{[1]},\dots, x^n_{[B]}, 
\hat{y}^n_{1[1]},\dots, \hat{y}^n_{1[B]}, x^n_{1[1]},\dots, x^n_{1[B]}, 
y^n_{[1]},\dots,y^n_{[B]}\right)\in \mathcal{T}_{\delta}\nonumber
\end{equation}
according to the stationary distribution of the Markov process
$G_{[l]}$ in (\ref{distr}).

The differences between our scheme and the CAF scheme are as
follows. At the transmitter side, in our scheme, the fresh message
$m_{[l]}$ is mapped into the codeword $x_{[l]}^n$ conditioned on the
codeword of the previous block $x_{[l-1]}^n$, while in the CAF scheme,
$m_{[l]}$ is mapped into $x_{[l]}^n$, which is generated independent
of $x_{[l-1]}^n$. At the relay side, in our scheme, the compressed
received signal $\hat{y}_{1[l-1]}^n$ is mapped into the codeword
$x_{1[l]}^n$, which is generated according to
$p(x_{1[l]}|\hat{y}_{1[l-1]})$, while in the CAF scheme, $x_{1[l]}^n$
is generated independent of $\hat{y}_{1[l-1]}^n$. The aim of our
design is to preserve the correlation built in the $(l-1)$st block in
the channel inputs of the $l$th block. At the decoding stage, we
perform joint decoding for the entire $B$ blocks after all of the $B$
blocks have been received, while in the CAF scheme, the decoding of
the message of the $(l-1)$st block is performed at the end of the
$l$th block.

\vspace*{0.1in}
\noindent 
{\bf Probability of error:} When $n$ is sufficiently large, the
probability of error can be made arbitrarily small when the following
conditions are satisfied.
\begin{enumerate}
\item For all $j$ such that $1\le j\le B-1$,
\begin{align}
\frac{1}{n}(B-j)\ln M+(B-j)&I(\hat{Y}_{1[l]};Y_{1[l]}|X_{1[l]}, 
X_{[l]})\nonumber\\
&< I(X_{[j]}^{[B-1]}, \hat{Y}_{1[j]}^{[B-1]}, X_{1[j+1]}^{[B]};Y_{[j]}^{[B]},
\hat{Y}_{1[B]},X_{[B]}| X_{[j-1]}, X_{1[j]})\label{cond4a}
\end{align}
\item For all $j,k$ such that $1\le j<k\le B-1$,
\begin{align}
\frac{1}{n}(B-j) &\ln M+(B-k)I(\hat{Y}_{1[l]};Y_{1[l]}|X_{1[l]}, 
X_{[l]})\nonumber\\
&< I(X_{[j]}^{[B-1]}, \hat{Y}_{1[k]}^{[B-1]}, X_{1[k+1]}^{[B]};Y_{[j]}^{[B]},
\hat{Y}_{1[B]},X_{1[B]},\hat{Y}_{1[j]}^{[k-1]},X_{1[j+1]}^{[k]}| X_{[j-1]}, 
X_{[j]})\label{cond4b}
\end{align}
\item For all $j,k$ such that $1\le k<j\le B-1$,
\begin{align}
(j-k)I(\hat{Y}_{1[l]};Y_{1[l]}|X_{1[l]}&,X_{[l]})+\frac{1}{n}(B-j)\ln M 
+(B-j)I(\hat{Y}_{1[l]};Y_{1[l]}|X_{1[l]}, X_{[l]})\nonumber\\
&<I(X_{[j]}^{[B-1]}, \hat{Y}_{1[k]}^{[B-1]}, X_{1[k+1]}^{[B]};Y_{[k]}^{[B]},
\hat{Y}_{1[B]},X_{[B]}|X_{[k]}^{[j-1]}, X_{1[k]})\label{cond4c}
\end{align}
\end{enumerate}
where the subscript $[l]$ on the left hand sides of (\ref{cond4a}), (\ref{cond4b}) and (\ref{cond4c}) indicates that the corresponding random variables belong to a generic sample $g_{[l]}$ of the underlying random process in (\ref{distr}).
The details of the calculation of the probability of error where these
conditions are obtained can be found in Appendix \ref{POE}. The derivation uses standard techniques from information theory, such as counting error events, etc.

In the above conditions, we used the notation $A_{[j]}^{[B]}$ as a
shorthand to denote the sequence of random variables $A_{[j]},
A_{[j+1]}, \dots, A_{[B]}$. Consequently, we note that the mutual
informations on the right hand sides of (\ref{cond4a}), (\ref{cond4b})
and (\ref{cond4c}) contain vectors of random variables whose lengths
go up to $B$, where $B$ is very large. In order to simplify the
conditions in (\ref{cond4a}), (\ref{cond4b}) and (\ref{cond4c}), we
lower bound the mutual information expressions on the right hand sides
of (\ref{cond4a}), (\ref{cond4b}) and (\ref{cond4c}) by those that
involve random variables that belong to up to three blocks. The detailed derivation of the
following lower bounding operation can be found in Appendix
\ref{LowB}.  %is
%omitted here due to space limitations. 
The derivation uses standard
techniques from information theory, such as the chain rule of mutual
information, and exploiting the Markov structure of the involved
random variables.
%This simplification is
%akin to the usual single-letterazation of $n$-letter mutual
%information expressions using chain rules and Markov properties in the
%usual converses \cite{Cover:1991}. 

\begin{enumerate}
\item For all $j$ such that $1\le j\le B-1$,
\begin{align}
(B-j) \left( \frac{1}{n}\ln M+I( \hat{Y}_{1[l]};Y_{1[l]}|X_{1[l]}, X_{[l]})\right)& 
\nonumber\\
< (B-j)I(Y_{[l]}&; X_{[l]},\hat{Y}_{1[l]}, X_{1[l]}|X_{[l-2]}, X_{1[l-1]}, 
Y_{[l-1]})\label{cond5a}
\end{align}
\item For all $j,k$ such that $1\le j<k\le B-1$,
\begin{align}
(k-j)\frac{1}{n}\ln M+(B-k)&\left(\frac{1}{n}\ln M+I(\hat{Y}_{1[l]};Y_{1[l]}|X_{1[l]}, X_{[l]})\right)
\nonumber\\
&< (k-j)I(X_{[l]}; Y_{[l]}, \hat{Y}_{1[l]}|X_{1[l]}, Y_{[l-1]}, 
\hat{Y}_{1[l-1]}, X_{1[l-1]}, X_{[l-2]})\nonumber\\
&\quad+(B-k)I(Y_{[l]};X_{[l]},\hat{Y}_{1[l]}, X_{1[l]}|X_{[l-2]}, X_{1[l-1]}, 
Y_{[l-1]})\label{cond5b}
\end{align}
\item For all $j,k$ such that $1\le k<j\le B-1$,
\begin{align}
(j-k)I(\hat{Y}_{1[l]};&Y_{1[l]}|X_{1[l]},X_{[l]})+(B-j)
\left(\frac{1}{n}\ln M +I(\hat{Y}_{1[l]};Y_{1[l]}|X_{1[l]}, X_{[l]})\right)\nonumber\\
&<(j-k)I(Y_{[l]};\hat{Y}_{1[l]}, X_{1[l]}|X_{[l]}, X_{[l-1]}, X_{1[l-1]}, 
Y_{[l-1]})\nonumber\\
&\quad +(B-j) I(Y_{[l]};X_{[l]},\hat{Y}_{1[l]}, X_{1[l]}|X_{[l-2]}, 
X_{1[l-1]}, Y_{[l-1]})\label{cond5c}
\end{align}
\end{enumerate}

We can further derive sufficient conditions for the above three
conditions in (\ref{cond5a}), (\ref{cond5b}) and (\ref{cond5c}) as
follows. We define the following quantities:
\begin{align}
C_1&\triangleq\frac{1}{n}\ln M+I(\hat{Y}_{1[l]};Y_{1[l]}|X_{1[l]}, X_{[l]})\\
C_2&\triangleq\frac{1}{n}\ln M\\
C_3&\triangleq I(\hat{Y}_{1[l]};Y_{1[l]}|X_{1[l]},X_{[l]})\\
D_1&\triangleq I(Y_{[l]};X_{[l]},\hat{Y}_{1[l]}, X_{1[l]}|X_{[l-2]}, 
X_{1[l-1]}, Y_{[l-1]})\\
D_2&\triangleq I(X_{[l]}; Y_{[l]}, \hat{Y}_{1[l]}|X_{1[l]}, Y_{[l-1]}, 
\hat{Y}_{1[l-1]}, X_{1[l-1]}, X_{[l-2]})\\
D_3&\triangleq I(Y_{[l]};\hat{Y}_{1[l]}, X_{1[l]}|X_{[l]}, X_{[l-1]}, 
X_{1[l-1]}, Y_{[l-1]})
\end{align}
Then, the sufficient conditions in (\ref{cond5a}), (\ref{cond5b}) and
(\ref{cond5c}) can also be written as,
\begin{enumerate}
\item For all $j$ such that $1\le j\le B-1$,
\begin{align}
(B-j)C_1 < (B-j)D_1\label{cond5d}
\end{align}
\item For all $j,k$ such that $1\le j<k\le B-1$,
\begin{align}
(k-j)C_2+(B-k)C_1 < (k-j)D_2+(B-k)D_1\label{cond5e}
\end{align}
\item For all $j,k$ such that $1\le k<j\le B-1$,
\begin{align}
(j-k)C_3+(B-j)C_1<(j-k)D_3+(B-j)D_1\label{cond5f}
\end{align}
\end{enumerate}
We note that the above conditions are implied by the following three
conditions,
\begin{align}
C_1&< D_1\\
C_2&< D_2\\
C_3&< D_3
\end{align}
or in other words, by,
\begin{align}
R-\eta\le\frac{1}{n}\ln M&< I(X_{[l]}; Y_{[l]}, \hat{Y}_{1[l]}|X_{1[l]}, Y_{[l-1]}, 
\hat{Y}_{1[l-1]}, X_{1[l-1]}, X_{[l-2]}) \label{cond5g}\\
I(\hat{Y}_{1[l]};Y_{1[l]}|X_{1[l]},X_{[l]})&<I(Y_{[l]};\hat{Y}_{1[l]}, 
X_{1[l]}|X_{[l]}, X_{[l-1]}, X_{1[l-1]}, Y_{[l-1]})\label{cond5h}\\
R-\eta+I(\hat{Y}_{1[l]};Y_{1[l]}|X_{1[l]}, X_{[l]})&< I(Y_{[l]};X_{[l]},
\hat{Y}_{1[l]}, X_{1[l]}|X_{[l-2]}, X_{1[l-1]}, Y_{[l-1]}) \label{cond5i}
\end{align}
The expressions in (\ref{cond5g}), (\ref{cond5h}) and (\ref{cond5i})
give sufficient conditions to be satisfied by the rate in order for
the probability of error to become arbitrarily close to zero. We note
that these conditions depend on variables used in three consecutive
blocks, $l$, $l-1$ and $l-2$. With this development, we obtain the
main result of our paper which is stated in the following theorem.

\begin{Theo}\label{Achrate}
The rate $R$ is achievable for the relay channel, if the following
conditions are satisfied
\begin{align}
R\le& I(Y,\hat{Y}_{1};X|  X_{1},\tilde{\hat{Y}}_{1},\tilde{Y}, \tilde{X}_{1},
\tilde{\tilde{X}})\label{cond6a}\\
I(\hat{Y}_1;Y_1|X_1,X)<& I(Y;\hat{Y}_{1},X_{1}|X, \tilde{Y},\tilde{X},  
\tilde{X}_{1})\label{cond6b}\\
R+I(\hat{Y}_1;Y_1|X_1,X)\le& I(Y;\hat{Y}_{1}, X_{1}, X_{}|\tilde{Y},
\tilde{X}_{1},\tilde{\tilde{X}})\label{cond6c}
\end{align}
where
%\begin{equation}
%\tilde{\tilde{X}}\longrightarrow(\tilde{X},\tilde{\hat{Y}}_1,\tilde{X}_1,
%\tilde{Y}, \tilde{Y}_1)\longrightarrow(X,\hat{Y}_1,X_1,Y, Y_1) 
%\label{cond6d}\end{equation}
%\begin{equation}
%p(x,\hat{y}_1,x_1,y,y_1,\tilde{x})=p(\tilde{x},\tilde{\hat{y}}_1,
%\tilde{x}_1, \tilde{y},\tilde{y}_1, \tilde{\tilde{x}}) \label{cond6e}\end{equation}
%\begin{align}
%p(x,&\hat{y}_1,x_1,y,y_1|\tilde{x},\tilde{\hat{y}}_1,\tilde{x}_1,
%\tilde{y},\tilde{y}_1)\nonumber\\
%&=p(x|\tilde{x})p(x_{1}|\tilde{\hat{y}}_{1})p(y_{1},y|x,x_{1})
%p(\hat{y}_1|y_1,x_1)
%\label{cond6f}
%\end{align}
\begin{align}
\tilde{\tilde{X}}\longrightarrow(\tilde{X},\tilde{\hat{Y}}_1,&\tilde{X}_1,
\tilde{Y}, \tilde{Y}_1)\longrightarrow(X,\hat{Y}_1,X_1,Y, Y_1) 
\label{cond6d}\\
p(x,\hat{y}_1,x_1,y,y_1,\tilde{x})&=p(\tilde{x},\tilde{\hat{y}}_1,
\tilde{x}_1, \tilde{y},\tilde{y}_1, \tilde{\tilde{x}}) \label{cond6e}\\
p(x,\hat{y}_1,x_1,y,y_1|\tilde{x},\tilde{\hat{y}}_1,\tilde{x}_1,
\tilde{y},\tilde{y}_1)&=p(x|\tilde{x})p(x_{1}|\tilde{\hat{y}}_{1})p(y_{1},y|x,x_{1})
p(\hat{y}_1|y_1,x_1)
\label{cond6f}
\end{align}
\end{Theo}

In the above theorem, the notations $\tilde{}$ and $\tilde{\tilde{}}$
are used to denote the signals belonging to the previous block and the
block before the previous block, respectively, with respect to a
reference block. Therefore, we see that the achievable rate in the
relay channel, using our proposed coding scheme, needs to satisfy
three conditions that involve mutual information expressions
calculated using eleven variables which satisfy the Markov chain
constraint in (\ref{cond6d}), the marginal distribution constraint in
(\ref{cond6e}), and the additional inter-block probability distribution constraint in (\ref{cond6f}).

In the next section, we will revisit the well-known CAF scheme
proposed in \cite{Cover:1979}. First, we will develop an equivalent
representation for the well-known representation of the achievable
rate in the CAF scheme. We will then show that the rates achievable by
the CAF scheme can be achieved with our proposed scheme by choosing a
certain special structure for the joint probability distribution of
the eleven random variables in Theorem~\ref{Achrate} while still
satisfying the three conditions in (\ref{cond6d}), (\ref{cond6e}) and
(\ref{cond6f}).
 
\section{Revisiting the Compress-And-Forward (CAF) Scheme}\label{revisit}
In \cite{Cover:1979}, the achievable rates for the CAF are
characterized as in the following theorem.
\begin{Theo}[\cite{Cover:1979}]\label{CAF}
The rate $R$ is achievable for the relay channel, if the following
conditions are satisfied
\begin{align}
R&\le I(X;Y,\hat{Y}_1|X_1)\label{cond1a}\\
I(Y_1;\hat{Y}_1|X_1,Y)&< I(X_1;Y) \label{cond1b}
\end{align}
where
\begin{equation}
p(x,x_1,y,y_1,\hat{y}_1)=p(x)p(x_1)p(y,y_1|x,x_1)p(\hat{y}_1|y_1,x_1)
\end{equation}
\end{Theo}
In the following theorem, we present three equivalent forms for
the rate achievable by  the CAF scheme.

\begin{Theo} \label{CAF-equiv}
The following three conditions are equivalent.
\begin{enumerate}
\item For some $p(x,x_1,y,y_1,\hat{y}_1)=p(x)p(x_1)p(y,y_1|x,x_1)p(\hat{y}_1|y_1,x_1)$
\begin{align}
R-I(X;\hat{Y}_1|X_1)&\le I(X;Y|\hat{Y}_1,X_1)\label{cond1c}\\
I(Y_1;\hat{Y}_1|X_1)&< I(\hat{Y}_1;Y|X_1)+I(X_1;Y)\label{cond1d}
\end{align}
\item For some $p(x,x_1,y,y_1,\hat{y}_1)=p(x)p(x_1)p(y,y_1|x,x_1)p(\hat{y}_1|y_1,x_1)$
\begin{align}
R-I(X;\hat{Y}_1|X_1)&\le I(X;Y|\hat{Y}_1,X_1)\label{cond2a}\\
R-I(X;\hat{Y}_1|X_1)+I(Y_1;\hat{Y}_1|X_1) 
&\le I(X, \hat{Y}_1;Y|X_1)+I(X_1;Y)\label{cond2b}
\end{align}
\item For some $p(x,x_1,y,y_1,\hat{y}_1)=p(x)p(x_1)p(y,y_1|x,x_1)p(\hat{y}_1|y_1,x_1)$
\begin{align}
R-I(X;\hat{Y}_1|X_1)&\le I(X;Y|\hat{Y}_1,X_1)\label{cond3a}\\
I(\hat{Y}_1;Y_1|X_1,X)&<
 I(\hat{Y}_1;Y|X_1,X)+I(X_1;Y|X)\label{cond3b}\\
R-I(X;\hat{Y}_1|X_1)+I(Y_1;\hat{Y}_1|X_1) 
&\le I(X, \hat{Y}_1;Y|X_1)+I(X_1;Y)\label{cond3c}
\end{align}
\end{enumerate}
\end{Theo}
The proof of the above theorem is given in Appendix
\ref{proof_CAF_equiv}.

We rewrite the final equivalent representation in (\ref{cond3a}),
(\ref{cond3b}) and (\ref{cond3c}) in the following more compact form
in order to compare the rates achievable with our proposed scheme and
the rates achievable with the CAF scheme in the next section.
\begin{align}
R&\le I(X;Y,\hat{Y}_1|X_1)\label{cond3d}\\
I(\hat{Y}_1;Y_1|X_1,X)&< I(\hat{Y}_1, X_1;Y|X)\label{cond3e}\\
R+I(Y_1;\hat{Y}_1|X_1,X)&\le I(X, \hat{Y}_1, X_1;Y)\label{cond3f}
\end{align}

\section{Comparison of the Achievable Rates with Our Scheme and 
with the CAF Scheme}

We note that the conditions on the achievable rates with our scheme
given in Theorem~\ref{Achrate}, i.e., (\ref{cond6a}), (\ref{cond6b}),
(\ref{cond6c}), are very similar to the final equivalent form for the
conditions on the achievable rates with the CAF scheme, i.e.,
(\ref{cond3d}), (\ref{cond3e}), (\ref{cond3f}), except for two
differences. First, the channel inputs of the transmitter and the
relay, i.e., $X$ and $X_1$, in our proposed scheme can be correlated,
while in the CAF scheme they are independent, and second, in our
scheme there are some extra random variables, which mutual information
expressions are conditioned on, e.g., $\tilde{X},\tilde{X}_{1},
\tilde{Y}, \tilde{\hat{Y}}_{1}, \tilde{\tilde{X}}$. These two
differences come from our coding scheme where we introduced
correlation between the channel inputs of the transmitter and the
relay in a block, and between the variables across the blocks. The
correlation between the channel inputs from the transmitter and the
relay in any block is an advantage, as for channels which favor
correlation, this translates into higher rates. However, the
correlation across the blocks is a disadvantage as it decreases the
efficiency of transmission, and therefore the achievable rates. In
fact, the price we pay for the correlation between the channel inputs
in any given block is precisely the correlation we have created across
the blocks. For a given correlation structure, it is not clear which
of these two opposite effects will overcome the other. That is, the
rate of our scheme for a certain correlated distribution may be lower
or higher than the rate of the CAF scheme. However, we note that the
CAF scheme can be viewed as a special case of our proposed scheme by
choosing an independent distribution, i.e., by choosing the following
conditional distribution in (\ref{cond6f})
\begin{equation}
p(x,\hat{y}_1,x_1,y, y_1|\tilde{x},\tilde{\hat{y}}_1,\tilde{x}_1,\tilde{y},\tilde{y}_1)
= p(x)p(x_{1})p(y_{1},y|x,x_{1})p(\hat{y}_1|x_1,y_1)
\end{equation}
In this case, the expressions in Theorem~\ref{Achrate}, i.e.,
(\ref{cond6a}), (\ref{cond6b}), (\ref{cond6c}), degenerate into the
third equivalent form for the CAF scheme in Theorem~\ref{CAF-equiv},
i.e., (\ref{cond3d}), (\ref{cond3e}), (\ref{cond3f}). The above
observation implies that the maximum achievable rate with our proposed
scheme over all possible distributions is not less than the achievable
rate of the CAF scheme. Thus, we can claim that this paper offers more
choices in the achievability scheme than the CAF scheme, and that
these choices may potentially yield larger achievable rates than those
offered by the CAF scheme. 

\appendix
\section{Appendix}
\subsection{Probability of Error Calculation}\label{POE}
The average probability of decoding error can be expressed as follows,
\begin{equation}
P_e=Pr(E_1\cup E_2)= Pr(E_1)+Pr(E_2\cap E_1^c)
\end{equation}
where
\begin{align}
E_1&\triangleq \left(x^n_{[1,\dots,B]},\hat{y}^n_{1[1,\dots, B]}, 
x^n_{1[1, \dots,B]}, y^n_{[1,\dots,B]}\right)\notin \mathcal{T}_{\delta}\\
E_2&\triangleq \bigcup_{\left(\bar{x}^n_{[1,\dots,B]},\bar{\hat{y}}^n_{1[1,
\dots, B-1]}\right)\neq \left(x^n_{[1,\dots,B]},\hat{y}^n_{1[1,\dots, B-1]}\right)} 
\left(\bar{x}^n_{[1,\dots,B]},\bar{\hat{y}}^n_{1[1,\dots, B]}, 
\bar{x}^n_{1[1, \dots,B]}, y^n_{[1,\dots,B]}\right)\in \mathcal{T}_{\delta}
\end{align}
where $(\bar{x}^n_{[1,\dots,B]},\bar{\hat{y}}^n_{1[1,\dots, B-1]},
\bar{x}^n_{1[2, \dots,B]})$ is another codeword that is generated
according to the rules of our scheme.

From  (\ref{distr}), we note the following Markov properties: 
\begin{enumerate}
\item conditioned on $(\hat{Y}_{1[l]}, X_{[l]}, X_{1[l]})$, $Y_{[l]}$
is independent of $G_{[\dots,l-1]}$ and  $G_{[l,\dots]}$; \label{Markov1}
\item conditioned on $(X_{[l-1]}, \hat{Y}_{1[l-1]})$,
$G_{[l,\dots]}$ is independent of
$G_{[\dots,l-1]}$.\label{Markov2}
\end{enumerate} 
Here, and in the sequel, subscript $[l]$ refers to a generic block within overall $B$ blocks. 

$Pr(E_1)$ can be upper bounded as follows:
\begin{align}
Pr(E_1)\le &\sum_{l=1}^B \left(Pr\left((x^n_{[l]},x^n_{1[l]}, y^n_{[l]},
y_{1[l]}^n,g^n_{[\dots,l-1]})\notin\mathcal{T}_{\delta}|
g^n_{[\dots,l-1]}\in\mathcal{T}_{\delta}\right)\right.\nonumber\\
&+\left.Pr\left((\hat{y}_{1[l]}^n,x^n_{[l]},x^n_{1[l]}, y^n_{[l]},
y_{1[l]}^n,g^n_{[\dots,l-1]})\notin
\mathcal{T}_{\delta}|(x^n_{[l]},x^n_{1[l]}, y^n_{[l]},
y_{1[l]}^n,g^n_{[\dots,l-1]})\in\mathcal{T}_{\delta}\right)\right)
\end{align}
From the way the code is generated, we have
\begin{equation}
Pr\left((x^n_{[l]},x^n_{1[l]}, y^n_{[l]},
y_{1[l]}^n,g^n_{[\dots,l-1]})\notin\mathcal{T}_{\delta}|
g^n_{[\dots,l-1]}\in\mathcal{T}_{\delta}\right)\le \epsilon
\end{equation}
The compression from $y_{1[l]}^n$ to $\hat{y}_{1[l]}^n$ is a
conditional version of a rate-distortion code. If $R'>
I(Y_1;\hat{Y}_1|X_1)$, then, when $n$ is sufficiently large, we have
\begin{equation}
Pr\left((\hat{y}_{1[l]}^n,x^n_{[l]},x^n_{1[l]}, y^n_{[l]},
y_{1[l]}^n,g^n_{[\dots,l-1]})\notin
\mathcal{T}_{\delta}|(x^n_{[l]},x^n_{1[l]}, y^n_{[l]},
y_{1[l]}^n,g^n_{[\dots,l-1]})\in\mathcal{T}_{\delta}\right)\le \epsilon
\end{equation}
Thus,
\begin{equation}
Pr(E_1)\le 2B\epsilon
\end{equation}

Now we switch to the error event $E_2$. 
\begin{align}
Pr&(E_2\cap E_1^c)\nonumber\\=&\sum_{\left(x^n_{[1,\dots,B]},\hat{y}^n_{1[1,\dots, B]}, 
x^n_{1[1, \dots,B]}, y^n_{[1,\dots,B]}\right)\in \mathcal{T}_{\delta}} p(x^n_{[1,\dots,B]},\hat{y}^n_{1[1,\dots, B]}, 
x^n_{1[1, \dots,B]}, y^n_{[1,\dots,B]}) \nonumber\\
&\qquad\qquad\qquad\qquad\qquad\qquad\qquad\times Pr\left(E_2|(x^n_{[1,\dots,B]},\hat{y}^n_{1[1,\dots, B]}, 
x^n_{1[1, \dots,B]}, y^n_{[1,\dots,B]}) \text{ sent}\right)\nonumber\\
\le & \max_{\left(x^n_{[1,\dots,B]},\hat{y}^n_{1[1,\dots, B]}, 
x^n_{1[1, \dots,B]}, y^n_{[1,\dots,B]}\right)\in \mathcal{T}_{\delta}}Pr\left(E_2|(x^n_{[1,\dots,B]},\hat{y}^n_{1[1,\dots, B]}, 
x^n_{1[1, \dots,B]}, y^n_{[1,\dots,B]}) \text{ sent}\right)
\end{align}

From our proposed coding scheme, we note that 
the codebooks at both transmitter and relay have tree structures with $B-1$ stages. A correct codeword $x^n_{[1,\dots,B-1]}$ can be viewed as a path in the tree-structured codebook at the transmitter. Similarly, for the codeword $\hat{y}^n_{1[1,\dots,B-1]}$ at the relay. An error occurs when we diverge from the correct path at a certain stage in the tree.
%we will not consider error events
%where the receiver decodes $x^n_{[j-1]}$ incorrectly but decodes
%$x^n_{[j]}$ correctly. Similarly, we will not consider error events
%where the receiver decodes $\hat{y}^n_{1[k-1]}$ incorrectly and
%$\hat{y}^n_{1[k]}$ correctly\footnote{Similar situations can be found
%in \cite{Cover:1980, Slepian:1973b}, where the error event of decoding
%the inner code (common information) incorrectly and the outer code
%(private information) correctly does not provide any constraints on
%the rate region.}. In other words, we will only consider the error
%events where the receiver decodes $x^n_{[1]},\dots,x^n_{[j-1]}$
%correctly and $x^n_{[j]},\dots, x^n_{[B-1]}$ incorrectly, and decodes
%$\hat{y}^n_{1[1]},\dots,\hat{y}^n_{1[k-1]}$ correctly and
%$\hat{y}^n_{1[k]},\dots, \hat{y}^n_{1[B-1]}$ incorrectly, for any
%$j,k\in \{1,\dots,B\}$. 
Thus, the error event $E_2$ can be decomposed
as
\begin{align}
&E_2=\bigcup_{\shortstack{${\scriptscriptstyle j=2,\dots,B-1}$\\$
{\scriptscriptstyle k=2,\dots,B-1}$}}\quad
\bigcup_{\shortstack{$
{\scriptscriptstyle \left(\bar{x}^n_{[1]},\dots,\bar{x}^n_{[j-1]},
\bar{\hat{y}}^n_{1[1]},\dots, \bar{\hat{y}}^n_{1[k-1]}\right)= 
\left(x^n_{[1]},\dots,x^n_{[j-1]},\hat{y}^n_{1[1]},\dots, \hat{y}^n_{1
[k-1]}\right)}$\\${\scriptscriptstyle \left(\bar{x}^n_{[j]},\bar{\hat{y}}^n_{1[k]}\right)\neq \left(x^n_{[j]},\hat{y}^n_{1[k]}
 \right) }$}}\nonumber\\
&\qquad\qquad\qquad\qquad
\left(\bar{x}^n_{[1]},\dots,\bar{x}^n_{[B]},\bar{\hat{y}}^n_{1
[1]},\dots, \bar{\hat{y}}^n_{1[B]}, \bar{x}^n_{1[1]}, \dots,\bar
{x}^n_{1[B]}, y^n_{[1]},\dots,y^n_{[B]}\right)\in \mathcal{T}_{\delta}
\end{align}
where each term in the union in the above equation represents the error event that results when we diverge from the correct paths at the $j$th stage at the transmitter and at the $k$th stage at the relay.

Let us define $\mathcal{F}_1$ to be the set consisting of all feasible
codeword pairs $(x_{[j]}^n, \hat{y}_{1[j]}^n)$ for the $j$th block for
a given $x_{[j-1]}^n$ and $x_{1[j]}^n$. Then, we have
\begin{align}
F_1\triangleq |\mathcal{F}_1|&\le M \exp(n(H(\hat{Y}_{1[j]}|X_{[j]}, 
X_{1[j]})+2\epsilon))\frac{L}{(1-\epsilon)\exp(n(H(\hat{Y}_{1[j]}| 
X_{1[j]})-2\epsilon))}\nonumber\\
&\le M \exp(n(H(\hat{Y}_{1[j]}|X_{[j]}, X_{1[j]})+2\epsilon))
\frac{\exp(n(I(\hat{Y}_{1[j]};Y_{1[j]}|X_{1[j]})+\epsilon))}{(1-\epsilon)
\exp(n(H(\hat{Y}_{1[j]}| X_{1[j]})-2\epsilon))}\nonumber\\
&\le M \exp(n(I(\hat{Y}_{1[j]};Y_{1[j]}|X_{1[j]}, X_{[j]})+6\epsilon))
\label{sizeKj}
\end{align}
We also define $\mathcal{F}_2$ to be the set consisting of all feasible
codewords $x_{[j]}^n$ for the $j$th block for a given
$x_{[j-1]}^n$. Then,
\begin{align}
F_2\triangleq |\mathcal{F}_2|&= M
\end{align}
Similarly, we define $\mathcal{F}_3$ to be the set consisting of all
feasible codewords $\hat{y}_{1[j]}^n$ for the $j$th block for a given
$x_{[j]}^n$ and $x_{1[j]}^n$. Then,
\begin{align}
F_3\triangleq |\mathcal{F}_3|&\le L\frac{\exp(n(H(\hat{Y}_{1[j]}|X_{1[j]},
X_{[j]})+2\epsilon))}{(1-\epsilon)\exp(n(H(\hat{Y}_{1[j]}|X_{1[j]})-2
\epsilon))}\nonumber\\
&\le \exp(n(I(\hat{Y}_{1[j]};Y_{1[j]}|X_{1[j]},X_{[j]})+6\epsilon))
\end{align}
We define the error event $E_{2jk}$ 
\begin{align}
E_{2jk}&\triangleq \bigcup_{\shortstack{$
{\scriptscriptstyle \left(\bar{x}^n_{[1]},\dots,\bar{x}^n_{[j-1]},
\bar{\hat{y}}^n_{1[1]},\dots, \bar{\hat{y}}^n_{1[k-1]}\right)= 
\left(x^n_{[1]},\dots,x^n_{[j-1]},\hat{y}^n_{1[1]},\dots, \hat{y}^n_{1
[k-1]}\right)}$\\${\scriptscriptstyle \left(\bar{x}^n_{[j]},\bar{\hat{y}}^n_{1[k]}\right)\neq \left(x^n_{[j]},\hat{y}^n_{1[k]}\right) }$}}\nonumber\\
&
\qquad\qquad\qquad\qquad\left(\bar{x}^n_{[1]},\dots,\bar{x}^n_{[B]},
\bar{\hat{y}}^n_{1[1]},\dots, \bar{\hat{y}}^n_{1[B]}, \bar{x}^n_{1[1]}, 
\dots,\bar{x}^n_{1[B]}, y^n_{[1]},\dots,y^n_{[B]}\right)\in 
\mathcal{T}_{\delta}
\end{align}
Then, we have
\begin{equation}
Pr(E_2\cap E_1^c)\le \sum_{j=2}^{B-1}\sum_{k=2}^{B-1}Pr(E_{2jk}\cap E_1^c)
\end{equation}
and
\begin{equation}
Pr(E_{2jk}\cap E_1^c) \le |\mathcal{A}_{jk}|\max_{(\bar{x}^n_{[1]},
\dots,\bar{x}^n_{[B-1]},\bar{\hat{y}}^n_{1[1]},\dots, 
\bar{\hat{y}}^n_{1[B-1]})\in\mathcal{A}_{jk}}P_1(\bar{x}^n_{[1]},
\dots,\bar{x}^n_{[B-1]},\bar{\hat{y}}^n_{1[1]},\dots, 
\bar{\hat{y}}^n_{1[B-1]})\label{errorjk}
\end{equation}
where
\begin{align}
&\mathcal{A}_{jk}
\triangleq \nonumber\\
&\left\{
\begin{array}{rcl}
\text{codeword }(\bar{x}^n_{[1]},\dots,\bar{x}^n_{[B-1]},
\bar{\hat{y}}^n_{1[1]},\dots, \bar{\hat{y}}^n_{1[B-1]}):&&\\
\left(\bar{x}^n_{[1]},\dots,\bar{x}^n_{[j-1]},\bar{\hat{y}}^n_{1[1]},
\dots, \bar{\hat{y}}^n_{1[k-1]}\right)&=& \left(x^n_{[1]},
\dots,x^n_{[j-1]},\hat{y}^n_{1[1]},\dots, \hat{y}^n_{1[k-1]}\right)\\ 
\left(\bar{x}^n_{[j]},\bar{\hat{y}}^n_{1[k]}\right)&\neq& \left(x^n_{[j]},\hat{y}^n_{1[k]}\right)
\end{array}\right\}\\
P_1&(\bar{x}^n_{[1]},\dots,\bar{x}^n_{[B-1]},\bar{\hat{y}}^n_{1[1]},
\dots, \bar{\hat{y}}^n_{1[B-1]})\nonumber\\
&\triangleq Pr((\bar{x}^n_{[1]},\dots,\bar{x}^n_{[B]},
\bar{\hat{y}}^n_{1[1]},\dots, \bar{\hat{y}}^n_{1[B]}, \bar{x}^n_{1[1]}, 
\dots,\bar{x}^n_{1[B]}, y^n_{[1]},\dots,y^n_{[B]})\in \mathcal{T}_{\delta})
\end{align}
given $\left(x^n_{[1]},\dots,x^n_{[B]},\hat{y}^n_{1[1]},\dots, \hat{y}^n_{1[B]}, x^n_{1[1]}, \dots,x^n_{1[B]}, y^n_{[1]},\dots,y^n_{[B]}\right)\in \mathcal{T}_{\delta}$.

In order to have the probability of such error events go to zero, we
need the following conditions to hold. %\footnote{A similar treatment for
%multiple-block joint decoding can be found in \cite{Kramer:1998}.}.

When $j=k$, from the structure of the block Markov code and
(\ref{sizeKj}), we have
\begin{equation}
|\mathcal{A}_{jk}|=
F_1^{B-j}\le M^{B-j} \exp(n(B-j)(I(\hat{Y}_{1[l]};Y_{1[l]}|X_{1[l]}, 
X_{[l]})+6\epsilon))\label{errorbegin}
\end{equation}
and 
\begin{align}
P_1(\bar{x}^n_{[1]},\dots,&\bar{x}^n_{[B-1]},\bar{\hat{y}}^n_{1[1]},
\dots, \bar{\hat{y}}^n_{1[B-1]})\nonumber\\
&\le\exp(n(H(X_{[j]}^{[B-1]}, \hat{Y}_{1[j]}^{[B-1]}, X_{1[j+1]}^{[B]}|
Y_{[j]}^{[B]}, \hat{Y}_{1[B]}, X_{[B]}, X_{[j-1]}, X_{1[j]})+2\epsilon))
\nonumber\\
&\quad\times\exp(-n(H(X_{[j]}^{[B-1]}, \hat{Y}_{1[j]}^{[B-1]}, 
X_{1[j+1]}^{[B]}|X_{[j-1]}, X_{1[j]})-2\epsilon))\nonumber\\
&=\exp(n(-I(X_{[j]}^{[B-1]}, \hat{Y}_{1[j]}^{[B-1]}, X_{1[j+1]}^{[B]};
Y_{[j]}^{[B]}, \hat{Y}_{1[B]}, X_{[B]}| X_{[j-1]}, X_{1[j]})+4\epsilon))
\end{align}

When $j< k$, we have
\begin{align}
|\mathcal{A}_{jk}|&=
F_2^{k-j}F_1^{B-k}\le M^{B-j}\exp(n(B-k)(I(\hat{Y}_{1[l]};Y_{1[l]}|X_{1[l]}, 
X_{[l]})+6\epsilon))
\end{align}
and 
\begin{align}
P_1&(\bar{x}^n_{[1]},\dots,\bar{x}^n_{[B-1]},\bar{\hat{y}}^n_{1[1]},
\dots, \bar{\hat{y}}^n_{1[B-1]})\nonumber\\
&\le\exp(n(H(X_{[j]}^{[B-1]}, \hat{Y}_{1[k]}^{[B-1]}, X_{1[k+1]}^{[B]}|
Y_{[j]}^{[B]},\hat{Y}_{1[B]},X_{[B]},\hat{Y}_{1[j]}^{[k-1]}, X_{[j-1]}, 
X_{1[j]}^{[k]})+2\epsilon))\nonumber\\
&\quad\times\exp(-n(H(X_{[j]}^{[B-1]}, \hat{Y}_{1[k]}^{[B-1]}, 
X_{1[k+1]}^{[B]}| X_{[j-1]}, X_{1[j]})-2\epsilon))\nonumber\\
&= \exp(n(-I(X_{[j]}^{[B-1]}, \hat{Y}_{1[k]}^{[B-1]}, X_{1[k+1]}^{[B]};
Y_{[j]}^{[B]},\hat{Y}_{1[B]},X_{[B]},\hat{Y}_{1[j]}^{[k-1]},
X_{1[j+1]}^{[k]}| X_{[j-1]}, X_{1[j]})+4\epsilon))
\end{align}

When $j> k$, we have
\begin{align}
|\mathcal{A}_{jk}|=F_3^{j-k}F_1^{B-j}&\le \exp(n(j-k)(I(\hat{Y}_{1[j]};
Y_{1[j]}|X_{1[j]},X_{[j]})+6\epsilon))\nonumber\\
&\quad\times M_l^{B-k} \exp(n(B-k)(I(\hat{Y}_{1[l]};Y_{1[l]}|X_{1[l]}, 
X_{[l]})+6\epsilon))
\end{align}
and
\begin{align}
P_1(\bar{x}^n_{[1]},\dots,&\bar{x}^n_{[B-1]},\bar{\hat{y}}^n_{1[1]},
\dots, \bar{\hat{y}}^n_{1[B-1]})\nonumber\\
&\le \exp(n(H(X_{[j]}^{[B-1]}, \hat{Y}_{1[k]}^{[B-1]}, X_{1[k+1]}^{[B]}|
Y_{[k]}^{[B]},\hat{Y}_{1[B]},X_{[B]},X_{k]}^{[j-1}, X_{1[k]})+2\epsilon))
\nonumber\\
&\quad\times \exp(-n(H(X_{[j]}^{[B-1]}, \hat{Y}_{1[k]}^{[B-1]}, 
X_{1[k+1]}^{[B]}|X_{[k]}^{[j-1]}, X_{1[k]})-2\epsilon))\nonumber\\
&=\exp(n(-I(X_{[j]}^{[B-1]}, \hat{Y}_{1[k]}^{[B-1]}, X_{1[k+1]}^{[B]};
Y_{[k]}^{[B]},\hat{Y}_{1[B]},X_{[B]}|X_{[k]}^{[j-1]}, X_{1[k]})+4\epsilon))\label{errorend}
\end{align}
Thus, when $n$ is sufficiently large, using (\ref{errorjk}) and (\ref{errorbegin}) through (\ref{errorend}), we have
\begin{equation}
Pr(E_{2jk}\cap E_1^c)\le \epsilon, \qquad j,k=2,\dots,B-1
\end{equation}
if the following conditions are satisfied:
\begin{enumerate}
\item For all $j$ such that $1\le j\le B-1$,
\begin{align}
\frac{1}{n}(B-j)\ln M+(B-j)&I(\hat{Y}_{1[l]};Y_{1[l]}|X_{1[l]}, 
X_{[l]})\nonumber\\
&< I(X_{[j]}^{[B-1]}, \hat{Y}_{1[j]}^{[B-1]}, X_{1[j+1]}^{[B]};Y_{[j]}^{[B]},
\hat{Y}_{1[B]},X_{[B]}| X_{[j-1]}, X_{1[j]})\label{cond4a2}
\end{align}
\item For all $j,k$ such that $1\le j<k\le B-1$,
\begin{align}
\frac{1}{n}(B-j) &\ln M+(B-k)I(\hat{Y}_{1[l]};Y_{1[l]}|X_{1[l]}, 
X_{[l]})\nonumber\\
&< I(X_{[j]}^{[B-1]}, \hat{Y}_{1[k]}^{[B-1]}, X_{1[k+1]}^{[B]};Y_{[j]}^{[B]},
\hat{Y}_{1[B]},X_{1[B]},\hat{Y}_{1[j]}^{[k-1]},X_{1[j+1]}^{[k]}| X_{[j-1]}, 
X_{[j]})\label{cond4b2}
\end{align}
\item For all $j,k$ such that $1\le k<j\le B-1$,
\begin{align}
(j-k)I(\hat{Y}_{1[l]};Y_{1[l]}|X_{1[l]}&,X_{[l]})+\frac{1}{n}(B-j)\ln M 
+(B-j)I(\hat{Y}_{1[l]};Y_{1[l]}|X_{1[l]}, X_{[l]})\nonumber\\
&<I(X_{[j]}^{[B-1]}, \hat{Y}_{1[k]}^{[B-1]}, X_{1[k+1]}^{[B]};Y_{[k]}^{[B]},
\hat{Y}_{1[B]},X_{[B]}|X_{[k]}^{[j-1]}, X_{1[k]})\label{cond4c2}
\end{align}
\end{enumerate}Therefore, we have
\begin{equation}
P_e=Pr(E_1)+Pr(E_2\cap E_1^c)\le (2B+B^2)\epsilon 
\end{equation}
When $n$ is sufficiently large, $(2B+B^2)\epsilon$ can be
made arbitrarily small.

\subsection{Lower Bounding the Mutual Informations
in (\ref{cond4a}), (\ref{cond4b}), (\ref{cond4c})}\label{LowB}
For the right hand side of (\ref{cond4a}), we have
\begin{align}
I(&X_{[j]}^{[B-1]}, \hat{Y}_{1[j]}^{[B-1]}, X_{1[j+1]}^{[B]};Y_{[j]}^{[B]},
\hat{Y}_{1[B]}, X_{[B]}| X_{[j-1]}, X_{1[j]})\nonumber\\
&\overset{\ref{reason11}}{=}\sum_{l=j}^{B-1}I(X_{[j]}^{[B-1]}, 
\hat{Y}_{1[j]}^{[B-1]}, X_{1[j+1]}^{[B]};Y_{[l]}| X_{[j-1]}, X_{1[j]}, 
Y_{[j]}^{[l-1]})\nonumber\\
&\quad +I(X_{[j]}^{[B-1]}, \hat{Y}_{1[j]}^{[B-1]}, X_{1[j+1]}^{[B]};Y_{[B]}, 
\hat{Y}_{1[B]},X_{[B]}| X_{[j-1]}, X_{1[j]}, Y_{[j]}^{[B-1]})\nonumber\\
&\overset{\ref{reason12}}{=}I(Y_{[j]};X_{[j]},\hat{Y}_{1[j]}| X_{1[j]}, 
X_{[j-1]})+\sum_{l=j+1}^{B-1} I(Y_{[l]};X_{[l]},\hat{Y}_{1[l]}, 
X_{1[l]}|X_{[j-1]}, X_{1[j]}, Y_{[j]}^{[l-1]})\nonumber\\
&\quad +I(Y_{[B]}, \hat{Y}_{1[B]},X_{[B]}; X_{1[B]}, X_{[B-1]}|X_{[j-1]}, 
X_{1[j]}, Y_{[j]}^{[B-1]})\nonumber\\
&\overset{\ref{reason13}}{=}\sum_{l=j+1}^{B-1} I(Y_{[l]};X_{[l]},
\hat{Y}_{1[l]}, X_{1[l]}|X_{[j-1]}, X_{1[j]}, Y_{[j]}^{[l-1]})
+ I(Y_{[B]};X_{[B]},\hat{Y}_{1[B]}| X_{1[B]}, X_{[B-1]})\nonumber\\
&\quad +I(Y_{[B]}, 
\hat{Y}_{1[B]},X_{[B]}; X_{1[B]}, X_{[B-1]}|X_{[j-1]}, X_{1[j]}, 
Y_{[j]}^{[B-1]})\nonumber\\
&\overset{\ref{reason14}}{\ge} \sum_{l=j+1}^{B-1} I(Y_{[l]};X_{[l]},
\hat{Y}_{1[l]}, X_{1[l]}|X_{[j-1]}, X_{1[j]}, Y_{[j]}^{[l-1]})\nonumber\\
&\quad + I(Y_{[B]};X_{[B]},\hat{Y}_{1[B]}| X_{1[B]}, X_{[B-1]}, X_{[j-1]}, 
X_{1[j]}, Y_{[j]}^{[B-1]})\nonumber\\
&\quad+I(Y_{[B]}; X_{1[B]}, X_{[B-1]}|X_{[j-1]}, X_{1[j]}, Y_{[j]}^{[B-1]})
\nonumber\\
&= \sum_{l=j+1}^{B-1} I(Y_{[l]};X_{[l]},\hat{Y}_{1[l]}, X_{1[l]}|X_{[j-1]}, 
X_{1[j]}, Y_{[j]}^{[l-1]})\nonumber\\
&\quad +I(Y_{[B]};X_{[B]},\hat{Y}_{1[B]}, X_{1[B]}, X_{[B-1]}| X_{[j-1]}, 
X_{1[j]}, Y_{[j]}^{[B-1]})\nonumber\\
&\overset{\ref{reason15}}{=}\sum_{l=j+1}^{B} I(Y_{[l]};X_{[l]},
\hat{Y}_{1[l]}, X_{1[l]}|X_{[j-1]}, X_{1[j]}, Y_{[j]}^{[l-1]})\nonumber\\
&\overset{\ref{reason16}}{\ge} (B-j)I(Y_{[l]};X_{[l]},\hat{Y}_{1[l]}, 
X_{1[l]}|X_{[l-2]}, X_{1[l-1]}, Y_{[l-1]})
\end{align}
where 
\begin{enumerate}
\item follows from the chain rule;\label{reason11}
\item because of Markov properties $1$ and $2$;\label{reason12}
\item because of the stationarity of the random process and the property
that conditioning reduces entropy;\label{reason13}
\item because of Markov property $2$;\label{reason14}
\item because of Markov property $1$;\label{reason15}
\item because of Markov property $2$ and the stationarity of the
random process.\label{reason16}
\end{enumerate}

For the right hand side of (\ref{cond4b}), we have 
\begin{align}
I(&X_{[j]}^{[B-1]}, \hat{Y}_{1[k]}^{[B-1]}, X_{1[k+1]}^{[B]};Y_{[j]}^{[B]},
\hat{Y}_{1[B]}, X_{[B]},\hat{Y}_{1[j]}^{[k-1]},X_{1[j+1]}^{[k]}| X_{[j-1]}, 
X_{1[j]})\nonumber\\
&\overset{\ref{reason21}}{=}I(X_{[j]}^{[B-1]}, \hat{Y}_{1[k]}^{[B-1]}, X_{1[k+1]}^{[B]};Y_{[j]}, 
\hat{Y}_{1[j]}|X_{[j-1]}, X_{1[j]})\nonumber\\
&\quad +\sum_{l=j+1}^{k-1}I(X_{[j]}^{[B-1]}, \hat{Y}_{1[k]}^{[B-1]}, 
X_{1[k+1]}^{[B]}; Y_{[l]}, \hat{Y}_{1[l]}, X_{1[l]}|X_{[j-1]},  
Y_{[j]}^{[l-1]}, \hat{Y}_{1[j]}^{[l-1]},X_{1[j]}^{[l-1]})\nonumber\\
&\quad+ I(X_{[j]}^{[B-1]}, \hat{Y}_{1[k]}^{[B-1]}, X_{1[k+1]}^{[B]}; 
Y_{[k]},X_{1[k]}|X_{[j-1]},  Y_{[j]}^{[k-1]}, \hat{Y}_{1[j]}^{[k-1]},
X_{1[j]}^{[k-1]})\nonumber\\
&\quad+\sum_{l=k+1}^{B-1} I(X_{[j]}^{[B-1]}, \hat{Y}_{1[k]}^{[B-1]}, 
X_{1[k+1]}^{[B]}; Y_{[l]}|X_{[j-1]},Y_{[j]}^{[l-1]},\hat{Y}_{1[j]}^{[k-1]},
X_{1[j]}^{[k]})\nonumber\\
&\quad+ I(X_{[j]}^{[B-1]}, \hat{Y}_{1[k]}^{[B-1]}, X_{1[k+1]}^{[B]}; 
Y_{[B]}, \hat{Y}_{1[B]},X_{[B]}|X_{[j-1]}, Y_{[j]}^{[B-1]},
\hat{Y}_{1[j]}^{[k-1]},X_{1[j]}^{[k]})\nonumber\\
&\overset{\ref{reason22}}{\ge} I(X_{[j]};Y_{[j]}, \hat{Y}_{1[j]}|X_{[j-1]}, X_{1[j]})+\sum_{l=j+1}^{k-1}
I(X_{[l]}; Y_{[l]}, \hat{Y}_{1[l]}|X_{[j-1]}, Y_{[j]}^{[l-1]}, 
\hat{Y}_{1[j]}^{[l-1]},X_{1[j]}^{[l]})\nonumber\\
&\quad+ I(X_{[k]}, \hat{Y}_{1[k]}; Y_{[k]}|X_{[j-1]},  Y_{[j]}^{[k-1]}, 
\hat{Y}_{1[j]}^{[k-1]},X_{1[j]}^{[k]})\nonumber\\
&\quad+\sum_{l=k+1}^{B-1} I(X_{[l]}, \hat{Y}_{1[l]}, X_{1[l]}; 
Y_{[l]}|X_{[j-1]},Y_{[j]}^{[l-1]},\hat{Y}_{1[j]}^{[k-1]},X_{1[j]}^{[k]})
\nonumber\\
&\quad+ I(X_{[B-1]}, X_{1[B]}; Y_{[B]}, \hat{Y}_{1[B]},X_{[B]}|X_{[j-1]}, 
Y_{[j]}^{[B-1]},\hat{Y}_{1[j]}^{[k-1]},X_{1[j]}^{[k]})\nonumber\\
&\overset{\ref{reason23}}{=}\sum_{l=j+1}^{k-1}I(X_{[l]}; Y_{[l]}, \hat{Y}_{1[l]}|X_{[j-1]}, 
Y_{[j]}^{[l-1]}, \hat{Y}_{1[j]}^{[l-1]},X_{1[j]}^{[l]})\nonumber\\
&\quad+\sum_{l=k+1}^{B-1} I(X_{[l]}, \hat{Y}_{1[l]}, X_{1[l]}; 
Y_{[l]}|X_{[j-1]},Y_{[j]}^{[l-1]},\hat{Y}_{1[j]}^{[k-1]},X_{1[j]}^{[k]})
\nonumber\\
&\quad+I(X_{[B]};Y_{[B]}, \hat{Y}_{1[B]}|X_{[B-1]}, X_{1[B]})\nonumber\\
&\quad+I(X_{[B]}, \hat{Y}_{1[B]}; Y_{[B]}|X_{[j-1+B-k]},  Y_{[j+B-k]}^{[B-1]}, 
\hat{Y}_{1[j+B-k]}^{[B-1]},X_{1[j+B-k]}^{[B]})\nonumber\\
&\quad+ I(X_{[B-1]}, X_{1[B]}; Y_{[B]}, \hat{Y}_{1[B]},X_{[B]}|X_{[j-1]}, 
Y_{[j]}^{[B-1]},\hat{Y}_{1[j]}^{[k-1]},X_{1[j]}^{[k]})\nonumber\\
&\overset{\ref{reason24}}{\ge} \sum_{l=j+1}^{k-1}I(X_{[l]}; Y_{[l]}, \hat{Y}_{1[l]}|X_{[j-1]}, 
Y_{[j]}^{[l-1]}, \hat{Y}_{1[j]}^{[l-1]},X_{1[j]}^{[l]})\nonumber\\
&\quad+\sum_{l=k+1}^{B-1} I(X_{[l]}, \hat{Y}_{1[l]}, X_{1[l]}; 
Y_{[l]}|X_{[j-1]},Y_{[j]}^{[l-1]},\hat{Y}_{1[j]}^{[k-1]},X_{1[j]}^{[k]})
\nonumber\\
&\quad+I(X_{[B]};Y_{[B]}, \hat{Y}_{1[B]}|X_{1[B]}, S)+I(X_{[B]}, 
\hat{Y}_{1[B]}, X_{1[B]}; Y_{[B]}| S)\nonumber\\
&\overset{\ref{reason25}}{\ge} (k-j)I(X_{[l]}; Y_{[l]}, \hat{Y}_{1[l]}|X_{1[l]}, Y_{[l-1]}, 
\hat{Y}_{1[l-1]}, X_{1[l-1]}, X_{[l-2]})\nonumber\\
&\quad+(B-k)I(Y_{[l]};X_{[l]},\hat{Y}_{1[l]}, X_{1[l]}|X_{[l-2]}, 
X_{1[l-1]}, Y_{[l-1]})
\end{align}
where 
\begin{equation}
S\triangleq (X_{[j-1+B-k]},  Y_{[j+B-k]}^{[B-1]}, \hat{Y}_{1[j+B-k]}^{[B-1]},
X_{1[j+B-k]}^{[B-1]}, X_{[j-1]}, Y_{[j]}^{[B-1]},\hat{Y}_{1[j]}^{[k-1]},
X_{1[j]}^{[k]})
\end{equation}
and
\begin{enumerate}
\item follows from the chain rule;\label{reason21}
\item because of Markov properties $1$ and $2$;\label{reason22}
\item because of the stationarity of the random process;\label{reason23}
\item because of the following derivation\label{reason24}
\begin{align}
I(&X_{[B]};Y_{[B]}, \hat{Y}_{1[B]}|X_{[B-1]}, X_{1[B]})\nonumber\\
&\quad+I(X_{[B]}, \hat{Y}_{1[B]}; Y_{[B]}|X_{[j-1+B-k]},  Y_{[j+B-k]}^{[B-1]}, 
\hat{Y}_{1[j+B-k]}^{[B-1]},X_{1[j+B-k]}^{[B]})\nonumber\\
&\quad+ I(X_{[B-1]}, X_{1[B]}; Y_{[B]}, \hat{Y}_{1[B]}, X_{[B]}|X_{[j-1]}, 
Y_{[j]}^{[B-1]},\hat{Y}_{1[j]}^{[k-1]},X_{1[j]}^{[k]})\nonumber\\
&\ge I(X_{[B]};Y_{[B]}, \hat{Y}_{1[B]}|X_{[B-1]}, X_{1[B]}, S)+I(X_{[B]}, 
\hat{Y}_{1[B]}; Y_{[B]}|X_{1[B]}, S)\nonumber\\
&\quad + I(X_{[B-1]}, X_{1[B]}; Y_{[B]}, \hat{Y}_{1[B]}|S)\nonumber\\
&\ge I(X_{[B]};Y_{[B]}, \hat{Y}_{1[B]}|X_{[B-1]}, X_{1[B]}, S)+I(X_{[B]}, 
\hat{Y}_{1[B]}; Y_{[B]}|X_{1[B]}, S)\nonumber\\
&\quad+I(X_{[B-1]}; Y_{[B]}, \hat{Y}_{1[B]}|X_{1[B]},S)+I(X_{1[B]};Y_{[B]}|S)
\nonumber\\
&= I(X_{[B]};Y_{[B]}, \hat{Y}_{1[B]}|X_{1[B]}, S)+I(X_{[B]}, \hat{Y}_{1[B]}, 
X_{1[B]}; Y_{[B]}| S)
\end{align}
\item because of Markov property $1$ and  $2$ and the stationarity of the
random process.\label{reason25}
\end{enumerate}
%The second inequality in the above derivation follows because of

For the right hand side of (\ref{cond4c}), we have 
\begin{align}
I(&X_{[j]}^{[B-1]}, \hat{Y}_{1[k]}^{[B-1]}, X_{1[k+1]}^{[B]};Y_{[k]}^{[B]},
\hat{Y}_{1[B]}, X_{[B]}|X_{[k]}^{[j-1]}, X_{1[k]})\nonumber\\
&\overset{\ref{reason31}}{=}\sum_{l=k}^{B-1}I(X_{[j]}^{[B-1]}, \hat{Y}_{1[k]}^{[B-1]}, X_{1[k+1]}^{[B]};
Y_{[l]}|X_{[k]}^{[j-1]}, X_{1[k]}, Y_{[k]}^{[l-1]})\nonumber\\
&\quad+I(X_{[j]}^{[B-1]}, \hat{Y}_{1[k]}^{[B-1]}, X_{1[k+1]}^{[B]};Y_{[B]}, 
\hat{Y}_{1[B]}, X_{[B]}|X_{[k]}^{[j-1]}, X_{1[k]}, Y_{[k]}^{[B-1]})\nonumber\\
&\overset{\ref{reason32}}{\ge}I(Y_{[k]};\hat{Y}_{1[k]}|X_{[k]}, X_{1[k]})+\sum_{l=k+1}^{j-1}I(Y_{[l]};
\hat{Y}_{1[l]}, X_{1[l]}|X_{[k]}^{[l]}, X_{1[k]}, Y_{[k]}^{[l-1]})\nonumber\\
&\quad+ I(Y_{[j]};X_{[j]}, \hat{Y}_{1[j]}, X_{1[j]}|X_{[k]}^{[j-1]}, 
X_{1[k]}, Y_{[k]}^{[j-1]})\nonumber\\
&\quad +\sum_{l=j+1}^{B-1}I(Y_{[l]};X_{[l]}, \hat{Y}_{1[l]}, 
X_{1[l]}|X_{[k]}^{[j-1]}, X_{1[k]}, Y_{[k]}^{[l-1]})\nonumber\\
&\quad + I(Y_{[B]}, \hat{Y}_{1[B]}, X_{[B]};X_{[j]}^{[B-1]}, 
X_{1[B]}|X_{[k]}^{[j-1]}, X_{1[k]}, Y_{[k]}^{[B-1]})\nonumber\\
&\overset{\ref{reason33}}{=}\sum_{l=k+1}^{j-1}I(Y_{[l]};\hat{Y}_{1[l]}, X_{1[l]}|X_{[k]}^{[l]}, 
X_{1[k]}, Y_{[k]}^{[l-1]})\nonumber\\
&\quad+\sum_{l=j+1}^{B-1}I(Y_{[l]};X_{[l]}, \hat{Y}_{1[l]}, 
X_{1[l]}|X_{[k]}^{[j-1]}, X_{1[k]}, Y_{[k]}^{[l-1]})
+I(Y_{[B]};\hat{Y}_{1[B]}|X_{[B]}, X_{1[B]})\nonumber\\
&\quad+ I(Y_{[B]};X_{[B]}, \hat{Y}_{1[B]}, X_{1[B]}|X_{[k+B-j]}^{[B-1]}, 
X_{1[k+B-j]}, Y_{[k+B-j]}^{[B-1]})\nonumber\\
&\quad + I(Y_{[B]}, \hat{Y}_{1[B]}, X_{[B]};X_{[j]}^{[B-1]}, 
X_{1[B]}|X_{[k]}^{[j-1]}, X_{1[k]}, Y_{[k]}^{[B-1]})\nonumber\\
& \overset{\ref{reason34}}{\ge} \sum_{l=k+1}^{j-1}I(Y_{[l]};\hat{Y}_{1[l]}, X_{1[l]}|X_{[j]}^{[l]}, 
X_{1[k]}, Y_{[k]}^{[l-1]})\nonumber\\
&\quad+\sum_{l=j+1}^{B-1}I(Y_{[l]};X_{[l]}, \hat{Y}_{1[l]}, 
X_{1[l]}|X_{[k]}^{[j-1]}, X_{1[k]}, Y_{[k]}^{[l-1]})\nonumber\\
&\quad +I(Y_{[B]};\hat{Y}_{1[B]}, X_{1[B]}|X_{[j]}^{[B]}, S')+
I(Y_{[B]};X_{[B]}, \hat{Y}_{1[B]}, X_{1[B]}| S')\nonumber\\
&\overset{\ref{reason35}}{\ge} (j-k)I(Y_{[l]};\hat{Y}_{1[l]}, X_{1[l]}|X_{[l]}, X_{[l-1]}, 
X_{1[l-1]}, Y_{[l-1]})\nonumber\\
&\quad +(B-j) I(Y_{[l]};X_{[l]},\hat{Y}_{1[l]}, X_{1[l]}|X_{[l-2]}, 
X_{1[l-1]}, Y_{[l-1]})
\end{align}
where 
\begin{equation}
S'\triangleq (X_{1[k+B-j]}, Y_{[k]}^{[B-1]}, X_{[k]}^{[j-1]}, X_{1[k]})
\end{equation}
and
\begin{enumerate}
\item follows from the chain rule;\label{reason31}
\item because of Markov properties $1$ and $2$;\label{reason32}
\item because of the stationarity of the random process;\label{reason33}
\item because of the following derivation\label{reason34}
\begin{align}
I(&Y_{[B]};\hat{Y}_{1[B]}|X_{[B]}, X_{1[B]})+ I(Y_{[B]};X_{[B]}, \hat{Y}_{1[B]}, X_{1[B]}|X_{[k+B-j]}^{[B-1]}, 
X_{1[k+B-j]}, Y_{[k+B-j]}^{[B-1]})\nonumber\\
&\quad + I(Y_{[B]}, \hat{Y}_{1[B]}, X_{[B]};X_{[j]}^{[B-1]}, 
X_{1[B]}|X_{[k]}^{[j-1]}, X_{1[k]}, Y_{[k]}^{[B-1]})\nonumber\\
&\ge I(Y_{[B]};\hat{Y}_{1[B]}|X_{[B]}, X_{1[B]}, S')+I(Y_{[B]};X_{[B]}, 
\hat{Y}_{1[B]}, X_{1[B]}|X_{[j]}^{[B-1]}, S')\nonumber\\
&\quad + I(Y_{[B]}, \hat{Y}_{1[B]}, X_{[B]};X_{[j]}^{[B-1]}, X_{1[B]}|S')
\nonumber\\
&= I(Y_{[B]};\hat{Y}_{1[B]}|X_{[B]}, X_{1[B]}, S')+I(Y_{[B]};X_{[B]}, 
\hat{Y}_{1[B]}, X_{1[B]}|X_{[j]}^{[B-1]}, S')\nonumber\\
&\quad + I(Y_{[B]}, \hat{Y}_{1[B]}, X_{[B]};X_{1[B]}| X_{[j]}^{[B-1]},S') 
+I(Y_{[B]}, \hat{Y}_{1[B]}, X_{[B]};X_{[j]}^{[B-1]}|S')\nonumber\\
&\ge I(Y_{[B]};\hat{Y}_{1[B]}|X_{[B]}, X_{1[B]}, X_{[j]}^{[B-1]}, S')+
I(Y_{[B]};X_{[B]}, \hat{Y}_{1[B]}, X_{1[B]}|X_{[j]}^{[B-1]}, S')\nonumber\\
&\quad + I(Y_{[B]};X_{1[B]}|X_{[B]}, X_{[j]}^{[B-1]},S') +I(Y_{[B]};
X_{[j]}^{[B-1]}|S')\nonumber\\
&=I(Y_{[B]};\hat{Y}_{1[B]}, X_{1[B]}|X_{[j]}^{[B]}, S')+I(Y_{[B]};X_{[B]}, 
\hat{Y}_{1[B]}, X_{1[B]}| S')
\end{align}
\item because of Markov property $1$ and  $2$ and the stationarity of the
random process.\label{reason35}
\end{enumerate}

\subsection{Proof of Theorem \ref{CAF-equiv}}\label{proof_CAF_equiv}
First, we note that condition $1$ is equivalent to the expression in
Theorem~\ref{CAF}. We also note that condition $2$ is seemingly weaker
than condition $1$ because (\ref{cond2b}) is implied by (\ref{cond1c})
and (\ref{cond1d}), and condition $3$ is seemingly stronger than
condition $2$ because condition $3$ consists of every element in
condition $2$ plus (\ref{cond3b}). Even though they seem different,
these three conditions are indeed equivalent. The equivalence of
conditions $2$ and $3$ is shown in \cite{Ahlswede:1983}. Here, we use
a similar proof technique to show the equivalence of conditions $1$
and $2$ as follows\footnote{A similar result is given in
\cite{Dabora:2006} by means of time-sharing.}. For a given
distribution $p(x,x_1,y,y_1,\hat{y}_1)$, condition $1$ is stronger
than condition $2$, which means that an arbitrary rate $R$ satisfying
condition $1$ will also satisfy condition $2$. Conversely, for a rate
$R$ satisfying condition $2$, if (\ref{cond1d}) is satisfied, then
condition $1$ is satisfied. If (\ref{cond1d}) is not satisfied, i.e.,
\begin{equation}
I(Y_1;\hat{Y}_1|X_1)\ge I(\hat{Y}_1;Y|X_1)+I(X_1;Y)
\end{equation}
we know that $R\in[0,R^*]$, where
\begin{align}
R^*-I(X;\hat{Y}_1|X_1)&\le I(X;Y|\hat{Y}_1,X_1)\label{cond2c}\\
R^*-I(X;\hat{Y}_1|X_1)+I(Y_1;\hat{Y}_1|X_1) 
&= I(X, \hat{Y}_1;Y|X_1)+I(X_1;Y)\label{cond2d}
\end{align}
That is, $R^*$ is defined such that (\ref{cond2b}) is satisfied with
equality.  We may rewrite (\ref{cond2c}) and (\ref{cond2d}) as
\begin{align}
R^*&\le I(X;Y|X_1)+I(X;\hat{Y}_1|Y,X_1)\\
R^*&= I(X,X_1;Y)-I(Y_1;\hat{Y}_1|X,X_1,Y)
\end{align}
We define a new random variable $\hat{Y}_1'$ such that $\hat{Y}_1'$
has the same marginal distribution as $\hat{Y}_1$ and
$\hat{Y}_1'\rightarrow \hat{Y}_1\rightarrow (Y_1, X,X_1,Y)$. Due to
the continuity of mutual information, there exists a choice of
$\hat{Y}_1'$ such that $I(X;\hat{Y}_1'|Y,X_1)=A$ for any $A\in[0,
I(X;\hat{Y}_1|Y,X_1)]$. If $R^*-I(X;Y|X_1)>0$, we choose $\hat{Y}_1'$
such that $R^*= I(X;Y|X_1)+I(X;\hat{Y}_1'|Y,X_1)$. We note that, in
this case, $I(Y_1;\hat{Y}_1|X,X_1,Y)\ge
I(Y_1;\hat{Y}_1'|X,X_1,Y)$. Thus,
\begin{align}
R^*&= I(X;Y|X_1)+I(X;\hat{Y}_1'|Y,X_1)\\
R^*&\le I(X,X_1;Y)-I(Y_1;\hat{Y}_1'|X,X_1,Y)
\end{align}
which means that $R^*$ satisfies condition $1$ with joint distribution
$p(x,x_1,y,y_1,\hat{y}_1')$ and so does any $R\le R^*$. If
$R^*-I(X;Y|X_1)\le0$, we choose $\hat{Y}_1'$ independent of
$(\hat{Y}_1, X,X_1, Y_1,Y)$. In this case,
\begin{align}
R^*&\le I(X;Y|X_1)+I(X;\hat{Y}_1'|Y,X_1)=I(X;Y|X_1)\\
0&=I(Y_1;\hat{Y}_1'|X_1)\le I(\hat{Y}_1';Y|X_1)+I(X_1;Y)
\end{align}
Therefore, in this case, $R^*$ satisfies condition $1$ with joint
distribution $p(x,x_1,y,y_1,\hat{y}_1')$ and so does any $R\le R^*$.

As we mentioned above the equivalence between condition $2$ and $3$ is
shown in \cite{Ahlswede:1983}. For completeness, we restate their
proof here as follows. For a given distribution
$p(x,x_1,y,y_1,\hat{y}_1)$, condition $3$ is stronger than condition
$2$, which means that an arbitrary rate $R$ satisfying condition $3$
will also satisfy condition $2$. Conversely, for a rate $R$ satisfying
condition $2$, if (\ref{cond3b}) is satisfied, then condition $3$ is
satisfied. If (\ref{cond3b}) is not satisfied, i.e., the following
inequalities are satisfied
\begin{align}
R-I(X;\hat{Y}_1|X_1)&\le I(X;Y|\hat{Y}_1,X_1)\\
I(\hat{Y}_1;Y_1|X_1,X)&\ge
 I(\hat{Y}_1;Y|X_1,X)+I(X_1;Y|X)\\
R-I(X;\hat{Y}_1|X_1)+I(Y_1;\hat{Y}_1|X_1) 
&\le I(X, \hat{Y}_1;Y|X_1)+I(X_1;Y)
\end{align}
then the following inequalities are satisfied also, since we simply
drop the first inequality,
\begin{align}
I(\hat{Y}_1;Y_1|X_1,X)&\ge
 I(\hat{Y}_1;Y|X_1,X)+I(X_1;Y|X)\label{cond3bm}\\
R-I(X;\hat{Y}_1|X_1)+I(Y_1;\hat{Y}_1|X_1) 
&\le I(X, \hat{Y}_1;Y|X_1)+I(X_1;Y)\label{cond3cm}
\end{align}
By combining (\ref{cond3bm}) and (\ref{cond3cm}), we have
\begin{align}
R\le & I(X;\hat{Y}_1|X_1)-I(Y_1;\hat{Y}_1|X_1)+I(\hat{Y}_1;Y_1|X_1,X)
\nonumber\\ 
& + I(X, \hat{Y}_1;Y|X_1)+I(X_1;Y)-I(\hat{Y}_1;Y|X_1,X)-I(X_1;Y|X)\nonumber\\
\le & I(X;Y|X_1)-(I(X_1;Y|X)-I(X_1;Y))	\nonumber\\
\le & I(X;Y|X_1)
\end{align}
which implies condition $3$, i.e., (\ref{cond3a}), (\ref{cond3b}) and
(\ref{cond3c}), with $\hat{Y}_1$ set to be a constant.

\bibliographystyle{unsrt}
\bibliography{refphd}
\end{document}